\documentclass[prc, reprint, amsmath, amssymb, superscriptaddress, aps, nofootinbib, showpacs]{revtex4-1}
\usepackage{graphicx}
\usepackage{epstopdf}

\usepackage{svg}
\usepackage{multirow}
\usepackage{epstopdf}
\usepackage[utf8]{inputenc}
\usepackage[english]{babel}
\usepackage{footnote}
\usepackage{color}
\usepackage{comment}

\usepackage{natbib}

\begin{document}

\title{First direct measurement of $^{48}$Ca single $\beta$-decay Q value with the TITAN Penning trap}

\author{S. Kakkar}
\affiliation{TRIUMF, 4004 Wesbrook Mall, Vancouver, British Columbia V6T 2A3, Canada}
\affiliation{Department of Physics and Astronomy, University of Manitoba, Winnipeg, Manitoba R3T 2N2, Canada}

\author{D. Perera}
\email{corresponding author: kadil1p@cmich.edu}
\affiliation{Department of Physics, Central Michigan University, Mount Pleasant, Michigan, 48859, USA}

\author{J. Ash}
\affiliation{TRIUMF, 4004 Wesbrook Mall, Vancouver, British Columbia V6T 2A3, Canada}

\author{B. Ashrafkhani}
\affiliation{TRIUMF, 4004 Wesbrook Mall, Vancouver, British Columbia V6T 2A3, Canada}
\affiliation{Department of Physics and Astronomy, University of Calgary, Calgary, Alberta T2N 1N4, Canada}

\author{J. D. Cardona}
\affiliation{TRIUMF, 4004 Wesbrook Mall, Vancouver, British Columbia V6T 2A3, Canada}
\affiliation{Department of Physics and Astronomy, University of Manitoba, Winnipeg, Manitoba R3T 2N2, Canada}

\author{C. Chambers}
\affiliation{TRIUMF, 4004 Wesbrook Mall, Vancouver, British Columbia V6T 2A3, Canada}

\author{A. Czihaly}
\affiliation{TRIUMF, 4004 Wesbrook Mall, Vancouver, British Columbia V6T 2A3, Canada}
\affiliation{Department of Physics and Astronomy, University of Victoria, Victoria, British Columbia V8P 5C2, Canada}

\author{G. Gwinner}
\affiliation{Department of Physics and Astronomy, University of Manitoba, Winnipeg, Manitoba R3T 2N2, Canada}

\author{Z. Hockenbery}
\affiliation{TRIUMF, 4004 Wesbrook Mall, Vancouver, British Columbia V6T 2A3, Canada}
\affiliation{Department of Physics, McGill University, Montreal, Quebec H3A 2T8, Canada}

\author{M. Horoi}
\affiliation{Department of Physics, Central Michigan University, Mount Pleasant, Michigan, 48859, USA}

\author{A. Kwiatkowski}
\affiliation{TRIUMF, 4004 Wesbrook Mall, Vancouver, British Columbia V6T 2A3, Canada}
\affiliation{Department of Physics and Astronomy, University of Victoria, Victoria, British Columbia V8P 5C2, Canada}

\author{E. Leistenschneider}
\affiliation{Nuclear Science Division, Lawrence Berkeley National Laboratory, 1 Cyclotron Road, Berkeley, 94720, CA, USA}

\author{E. M. Lykiardopoulou}
\affiliation{TRIUMF, 4004 Wesbrook Mall, Vancouver, British Columbia V6T 2A3, Canada}
\affiliation{Department of Physics and Astronomy, University of British Columbia, Vancouver, British Columbia V6T 1Z1, Canada}

\author{F. Maldonado Mill\'{a}n}
\affiliation{TRIUMF, 4004 Wesbrook Mall, Vancouver, British Columbia V6T 2A3, Canada}

\author{A. Mollaebrahimi}
\affiliation{TRIUMF, 4004 Wesbrook Mall, Vancouver, British Columbia V6T 2A3, Canada}
\affiliation{Physikalisches Institut, Justus-Liebig-Universit\"{a}t Gie\ss en, 35392 Gie\ss en, Germany}

\author{S. Paul}
\affiliation{TRIUMF, 4004 Wesbrook Mall, Vancouver, British Columbia V6T 2A3, Canada}

\author{W. S. Porter}
\affiliation{Department of Physics and Astronomy, University of Notre Dame, Notre Dame, Indiana, 46556, USA}

\author{D. Ray}
\affiliation{TRIUMF, 4004 Wesbrook Mall, Vancouver, British Columbia V6T 2A3, Canada}

\author{M. Redshaw}
\affiliation{Department of Physics, Central Michigan University, Mount Pleasant, Michigan, 48859, USA}

\author{R. Ringle}
\affiliation{Facility for Rare Isotope Beams, East Lansing, MI 48824, USA}

\author{C. Walls}
\affiliation{TRIUMF, 4004 Wesbrook Mall, Vancouver, British Columbia V6T 2A3, Canada}
\affiliation{Department of Physics and Astronomy, University of Manitoba, Winnipeg, Manitoba R3T 2N2, Canada}

\author{A. Weaver}
\affiliation{TRIUMF, 4004 Wesbrook Mall, Vancouver, British Columbia V6T 2A3, Canada}

\date{\today}%

\begin{abstract}
\begin{description}
\item[Background]
Neutrinoless double $\beta$-decay (0$\nu\beta\beta$), if observed, would provide unequivocal evidence of physics beyond the Standard Model. $^{48}$Ca is an interesting candidate system to study because it has the largest $Q_{\beta\beta}$ value among all $\beta\beta$ transitions and is also unstable against single $\beta$-decay. The observation of both $\beta$ and $\beta\beta$-decay in the same isotope would provide a unique opportunity to benchmark theoretical calculations of $\beta$ and $\beta\beta$-decay matrix elements and could provide insight on the quenching of the axial vector coupling constant, $g_{A}$.

\item[Purpose]
Perform a precise measurement of the mass of $^{48}$Sc to enable a new determination of the $^{48}$Ca -- $^{48}$Sc single $\beta$-decay Q value and perform an updated determination of the $^{48}$Ca single $\beta$-decay lifetime to compare with experimental lower limits.

\item[Method]
The TITAN Penning trap mass spectrometer at the TRIUMF facility was utilized for a high-precision measurement of the $^{48}$Ca single $\beta$-decay Q value. This was achieved through cyclotron frequency ratio measurements of $^{48}$Ca$^{+}$/$^{48}$Sc$^{+}$ and $^{48}$Sc$^{+}$/$^{48}$Ti$^{+}$ using the Time-of-Flight Ion Cyclotron Resonance (TOF-ICR) technique. 

\item[Results]
The $^{48}$Ca $\beta$-decay Q value was determined to be $Q_{\beta}$($^{48}$Ca) = 279.14(50) keV, a factor of 10 more precise than the previous value given in the 2020 Atomic Mass Evaluation [Wang, \textit{et al.}, Chin. Phys. C \textbf{45}, 030003 (2021)]. This Q value was used to determine the $^{48}$Ca single $\beta$-decay partial half-life, with the result $T_{1/2}^{\beta}$($^{48}$Ca) = 5.09(5) $\times$ 10$^{20} (g_A^{-2})$ y. 

\item[Conclusion]
Our $^{48}$Ca single $\beta$-decay half-life was determined to a precision of 1\%, a factor of 30 improvement compared to calculations using the previous Q value. Our result is marginally closer to the experimental lower limit $T_{1/2}^{\beta}$($^{48}$Ca) $>$ 1.1 $\times 10^{20}$ y, but still a factor 5 longer. It is also a factor of 10 longer than the observed two-neutrino $\beta\beta$-decay mode with $T_{1/2}^{2\nu\beta\beta}$($^{48}$Ca) = 5.96$^{+1.39}_{-1.08}\times 10^{19}$ y.
Hence, it could be possible to observe $^{48}$Ca single $\beta$-decay in future experiments, strengthening the potential importance of $^{48}$Ca to benchmark nuclear structure and $\beta\beta$-decay studies.

\end{description}
\end{abstract}

\maketitle

\section{Introduction}
\begin{figure}[t]
    \includegraphics[width=0.9\columnwidth]{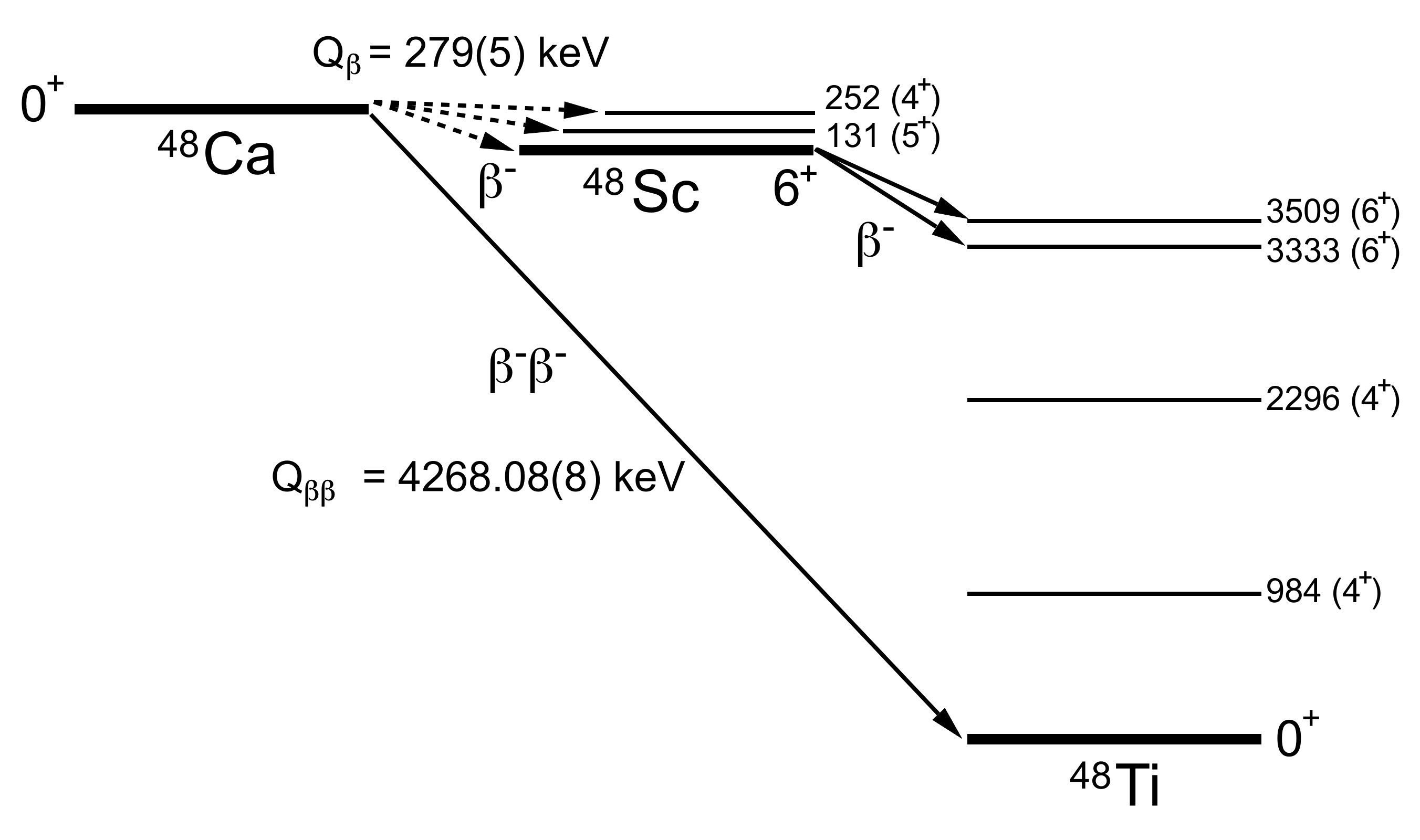}
    \caption{Decay scheme for $^{48}$Ca $\beta\beta$-decay, as yet unobserved $^{48}$Ca single-$\beta$-decay, and $^{48}$Sc $\beta$-decay. The ground-state to ground-state Q values are obtained using data from the AME2020~\cite{Wang_2021}. Energy levels are given in units of keV.}
    \label{fig:48Ca_Decay_Scheme}
\end{figure}

Searches for neutrinoless double $\beta$-decay (0$\nu\beta\beta$) provide the exciting possibility of discovering physics beyond the Standard Model. If observed, 0$\nu\beta\beta$ would indicate that the neutrino is a Majorana particle and that lepton number is not conserved. It would also enable a determination of the effective Majorana neutrino mass. Among the 35 naturally occurring primordial isotopes that are $\beta\beta$-decay candidates (where single $\beta$-decay is energetically or otherwise forbidden), 11 are considered the most prominent for 0$\nu\beta\beta$ searches because they have Q values, $Q_{\beta\beta}$ $>$ 2 MeV, which results in a higher expected decay rate. $^{48}$Ca stands out among these as the isotope with the highest Q value, $Q_{\beta\beta}$($^{48}$Ca) = 4.3 MeV. It is also unstable against single $\beta$-decay, a property shared with only one other isotope, $^{96}$Zr. However, for both isotopes, single $\beta$-decay is highly forbidden due to the large angular momentum difference between initial and final nuclear states: $\Delta J$= 4, 5 or 6 (see the decay scheme in Fig. \ref{fig:48Ca_Decay_Scheme}).

In $^{48}$Ca, as with a number of other $\beta\beta$-decay candidates~\cite{Barabash2020,Barabash2023,Pritychenko2025}, the rare but Standard Model allowed $2\nu\beta\beta$ decay mode has been observed~\cite{Balysh1996_48Ca,Brudanin2000_48Ca,Arnold2016_48Ca} with an evaluated half-life $T_{1/2}^{2\nu\beta\beta}$($^{48}$Ca) = 5.96$^{+1.39}_{-1.08}\times 10^{19}$ y~\cite{Pritychenko2025}. 
The single $\beta$-decay mode on the other hand has not been detected~\cite{Alburger1985_48Ca,Bakalyarov2002_48Ca_Full,Bernabei2002_48Ca}, but an experimental lower limit of $T_{1/2}^{\beta}$($^{48}$Ca) $>$ 1.1 $\times 10^{20}$ y has been established~\cite{Bakalyarov2002_48Ca_Full}.
Shell model calculations that determine the $^{48}$Ca $\beta$-decay half-life~\cite{Warburton1985_48Ca,Aunola1999_48Ca,Haaranen2014,Kostensal2020_48Ca} show that it is dominated by the 4$^{\mathrm{th}}$ forbidden decay to the 131 keV, 5$^{+}$ state in $^{48}$Sc, with $T_{1/2}^{\beta}$($^{48}$Ca)  = 5.3 $\times 10^{20}$ y~\cite{Haaranen2014}, a factor of 5 larger than the current experimental lower limit~\cite{Bakalyarov2002_48Ca_Full}. Nevertheless, the question remains: could $^{48}$Ca single $\beta$-decay be experimentally observed? If so, comparison of experimental and calculated $\beta$ and $\beta\beta$-decay rates would provide a unique opportunity to benchmark theoretical calculations of nuclear matrix elements, and could provide insight into quenching of the axial vector coupling constant, $g_{A}$~\cite{Kostensal2020_48Ca}.

Currently, the $\approx$30 \% uncertainty in the calculated single $\beta$-decay rate is dominated by uncertainty in the $^{48}$Ca -- $^{48}$Sc $\beta$-decay Q value. The Q value, obtained from atomic mass data from the most recent atomic mass evaluation (AME2020)~\cite{Wang_2021} is $Q_{\beta}$ = 279(5) keV. The 5 keV uncertainty in the Q value is primarily due to the uncertainty in the mass of $^{48}$Sc, which is currently determined from $^{48}$Ca($p,n$)$^{48}$Sc reaction data~\cite{McMurray1967,McDonald1968} and a measurement of the $^{48}$Sc $\beta$-decay spectrum end-point energy~\cite{vanNooijen1957}. Hence, there is an urgent need to both reduce the uncertainty in the Q value, and to check its accuracy to enable a new determination of the $^{48}$Ca single $\beta$-decay rate and evaluate the feasibility of detecting $^{48}$Ca single $\beta$-decay in future experiments. In this paper, we present the first direct measurement of the $^{48}$Ca single $\beta$-decay Q value via Penning trap mass spectrometry with the TITAN Penning trap at TRIUMF. We also discuss updated calculations of the $^{48}$Ca single $\beta$-decay rate and the implications for observing this highly forbidden decay.
 
\section{Experimental Description}\label{Section:Exp}

\begin{figure}[ht]
\includegraphics[width=0.9\columnwidth]{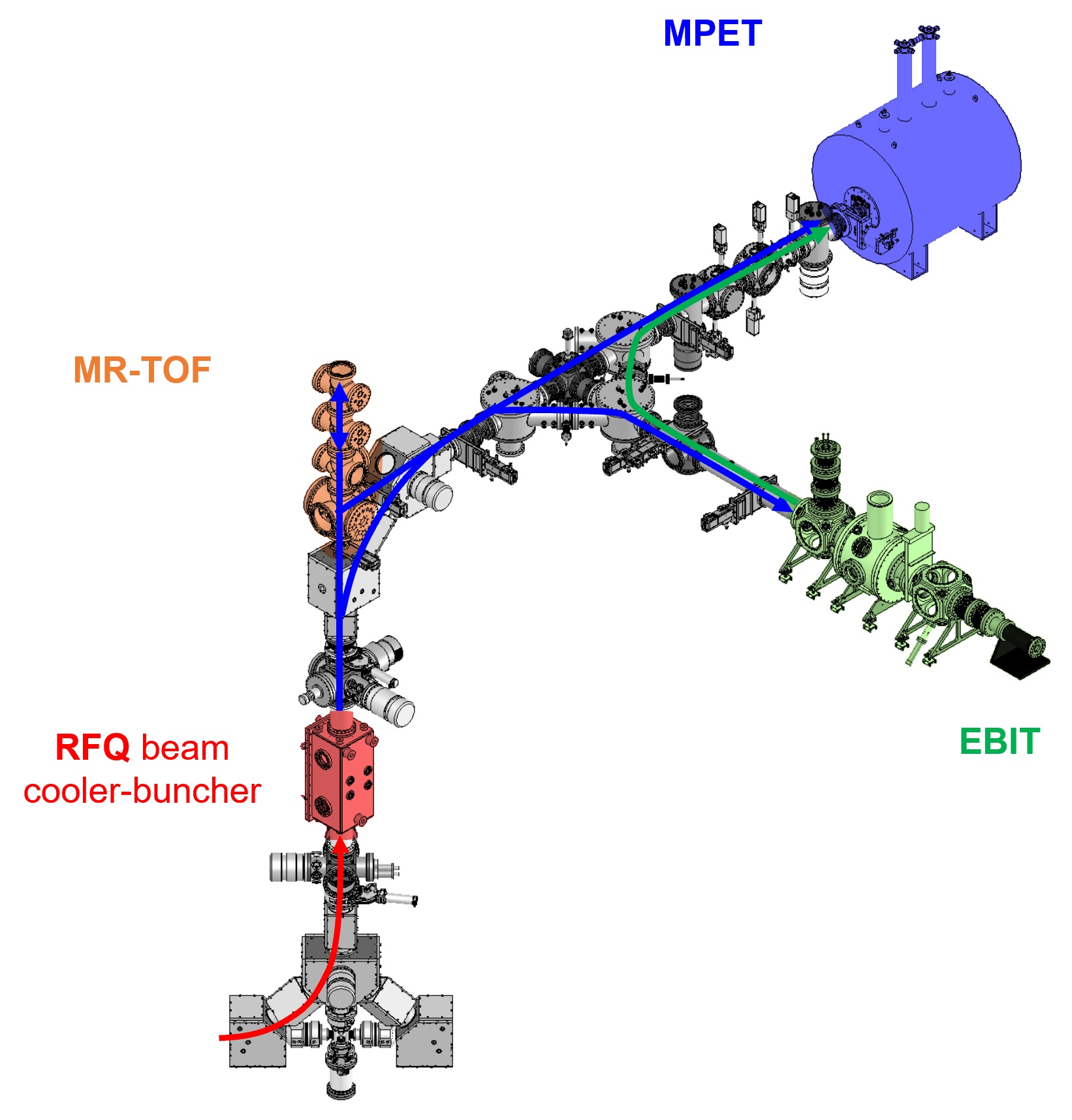}
\caption{Schematic of the TITAN-TRIUMF beamline. Ions are first delivered to the TITAN RFQ beam cooler and buncher. The bunched ion beam is then sent either to the MR-TOF for identification of isobaric species, or to the measurement Penning trap (MPET) where the mass measurement is performed with the TOF-ICR technique. In this work the EBIT was not used.\label{fig:TITAN beam line}} 
\end{figure}

In this work, the TITAN Penning trap facility at TRIUMF (see Fig.~\ref{fig:TITAN beam line}) was used to perform high-precision cyclotron frequency ratio measurements between singly-charged $^{48}$Ca, $^{48}$Sc, and $^{48}$Ti ions. 


The ions of interest were provided by the TRIUMF Isotope Separation and ACceleration (ISAC) facility~\cite{Ball2016_TRIUMF} that produces rare isotopes via the isotope separation on-line (ISOL) technique. In this experiment, the 44 h half-life isotope $^{48}$Sc was produced by a 70 $\mu$A, 520 MeV proton beam impinged on a low-power tantalum target, used in combination with an ion guide laser ion source (IG-LIS)~\cite{Mostamand2020_IGLIS}. The production method also generated surface ionized $^{48}$Ca$^{+}$ and $^{48}$Ti$^{+}$ ions that were used in the measurement, along with the 6.8 s half-life isotope $^{48}$K$^{+}$.

An initial purification step was performed using a mass separator with a resolving power $R \approx 2500$ that enabled the selection of the $A = 48$ isobars and removal of molecular contaminants. Next, ions were delivered to the TITAN facility as a continuous 20~keV ion beam that first entered the radio-frequency quadrupole (RFQ) cooler and buncher~\cite{TITANrfq}. The cooler-buncher is a linear Paul trap filled with helium buffer gas, in which ions are cooled via collisions with the gas, reducing the transverse emittance and energy spread of the beam. The accumulated ions were extracted as low-emittance bunches, accelerated to 2.2 keV, and delivered to either the Multi-Reflection Time-Of-Flight Mass Spectrometer (MR-TOF)~\cite{Jesch2014_MRTOF,Reiter2021_MRTOF} or the Measurement Penning Trap (MPET)~\cite{Brodeur2012}.

In the MR-TOF, ions are confined between two electrostatic mirrors and undergo multiple reflections, extending their flight path within a compact geometry. Since time-of-flight scales as $t \propto \sqrt{m/q}$, ions with different $m/q$ values separate in time as they complete successive reflections. The TITAN MR-TOF has been used for mass measurements and mass separation, achieving resolving powers of up to $\sim$600,000~\cite{Porter2022} and 250,000~\cite{Reiter2021_MRTOF}, respectively. In this work the MR-TOF was used to identify beam constituents and optimize delivery from ISAC, ensuring optimal conditions for precision mass measurements with MPET.
%
%
Downstream from the MR-TOF is the Electron Beam Ion Trap (EBIT), which is used for charge breeding, but this step was not used in this experiment. 


In the Penning trap, ions are confined by the superposition of a strong homogeneous magnetic field, $\vec{B} = B\hat{z}$, that provides radial confinement, and a quadrupolar electrostatic potential, that  confines the ions in the axial direction with a simple-harmonic motion at the axial frequency $\nu_{z}$. In Penning trap mass spectrometry, the goal is to determine the mass of an ion via its free-space cyclotron frequency~\cite{Brown1986,Blaum2006},
\begin{equation}\label{Eqn_fc}
\nu_c = \frac{qB}{2\pi m}.
\end{equation}
\noindent However, the electric field has the effect of modifying an ion's cyclotron motion, so that inside the trap it revolves at the reduced-cyclotron frequency, $\nu_{+}$. The electric field also introduces a second radial motion, the magnetron motion, at frequency $\nu_{-}$, that results from the $\vec{E} \times \vec{B}$ drift experienced by an ion in the trap. The free-space cyclotron frequency of Eqn.~(\ref{Eqn_fc}) can be recovered by combinations of the trap frequencies, $\nu_{+}$, $\nu_{-}$, and $\nu_{z}$. Of particular relevance for this work is the relationship
\begin{equation}\label{Eqn_f+-}
\nu_c = \nu_+ + \nu_-,
\end{equation}
\noindent that holds true for an ideal Penning trap, and can be considered exact to the level of precision of the measurements presented here~\cite{Gabrielse2009_IJMS,Gabrielse2009_PRL}.
At TITAN, the MPET operates in a $B$ = 3.7~T field, provided by a superconducting solenoidal magnet, resulting in a cyclotron frequency for singly-charged $A = 48$ ions of approximately 1.2 MHz. The TITAN MPET was recently upgraded to the \textit{CryoMET} design for operation at cryogenic temperatures, and the work presented here represents the first online measurement after this upgrade~\cite{Leistenschneider_PhD,Lykiardopoulou_PhD,Kakkar_PhD}.

\begin{figure}[t]
\includegraphics[width=0.9\columnwidth]{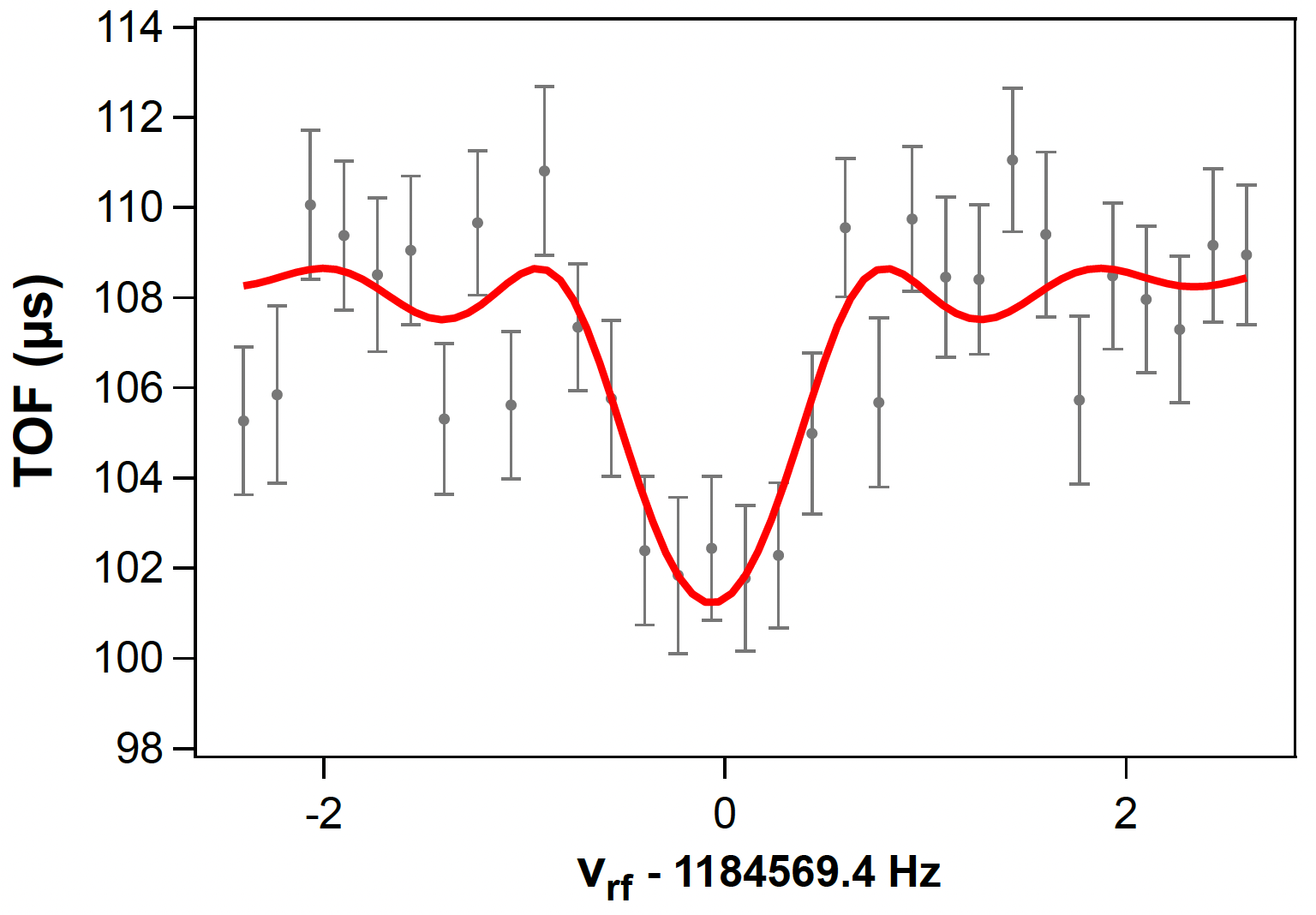}
\caption{(Color online) Time-of-flight single resonance curve for $^{48}$Sc$^{+}$ with an excitation time of 1 s. The solid red line is the fit of the theoretical line shape to the experimental data~\cite{Konig1995}.}. 
\label{fig:Single_Res}
\end{figure}
\begin{figure}[b]
\includegraphics[width=0.9\columnwidth]{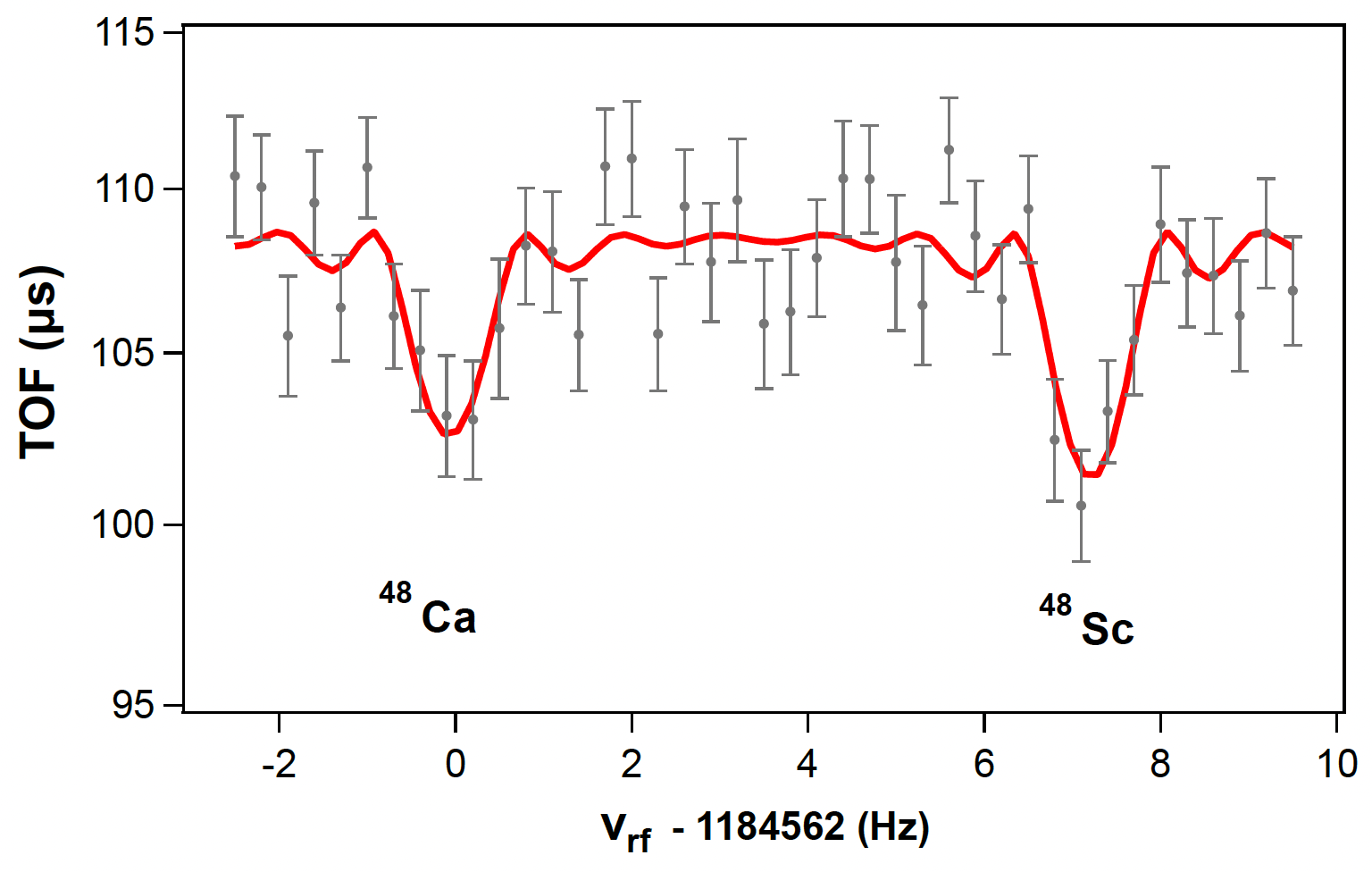}
\caption{(Color online) Time-of-flight simultaneous double resonance curve for $^{48}$Ca$^{+}$ and $^{48}$Sc$^{+}$ with an excitation time of 1 s. The solid red line is the fit of the theoretical line shape to the experimental data~\cite{Konig1995}.}. 
\label{fig:Double_Res}
\end{figure}

In the MPET, the cyclotron frequency was determined using the time-of-flight ion cyclotron resonance (TOF-ICR) technique~\cite{Graeff1980,Konig1995}. Ions were first prepared with initial magnetron motion by injecting them off-center from the trap axis using a Lorentz steerer~\cite{Ringle2007_Lor}. 
A ``dipole cleaning'' radio frequency (rf) drive was then applied at the trap cyclotron frequency corresponding to the other ions that were delivered in the beam from ISAC to remove them from the center of the trap and avoid potential frequency shifts due to Coulomb interactions.
Ions were then subjected to a quadrupolar rf excitation at $\nu_{\mathrm{rf}} \approx \nu_{+} + \nu_{-}$, which couples the ions' radial motions, resulting in an increase in radial energy when magnetron motion is converted into cyclotron motion. Ions were subsequently ejected from the trap toward a microchannel plate detector (MCP) and their time-of-flight (TOF) was recorded. 
%
%
The TOF is minimized for ions with a larger radial energy, which occurs when $\nu_{\mathrm{rf}} = \nu_{c}$. Data is acquired by scanning $\nu_{\mathrm{rf}}$ over a frequency range close to $\nu_{c}$ for subsequent bunches of ions and creating a TOF vs. $\nu_{\mathrm{rf}}$ spectrum. A fit of the theoretical line shape~\cite{Konig1995} to the data can then be performed to extract $\nu_{c}$, as shown in Figs.~\ref{fig:Single_Res} and \ref{fig:Double_Res}. During the course of the experiment, we took data with rf quadrupole excitation times of 0.5 s, 1.0 s and 2.0 s.


In order to obtain cyclotron frequency ratios of ions and account for linear magnetic field drifts, we followed the standard procedure of alternating between cyclotron frequency measurements, as described above, on pairs of ions. Additionally, for $^{48}$Ca and $^{48}$Sc we obtained simultaneous double resonance TOF measurements. This was possible because $^{48}$Ca and $^{48}$Sc have a fractional mass difference of $6 \times 10^{-6}$, resulting in a cyclotron frequency difference in the TITAN MPET of only 7 Hz. Therefore, with $^{48}$Ca$^{+}$ and $^{48}$Sc$^{+}$ being delivered at approximately equal rates, we performed a TOF-ICR measurement that covered $\nu_{c}$ for both ions. An example of such a resonance is shown in Fig.~\ref{fig:Double_Res}.

\section{Data and Analysis}

Our complete data set consisted of 13 alternating cyclotron frequency ratios of $^{48}$Ca$^+$/$^{48}$Sc$^+$, \color{black}{7 }\color{black} of $^{48}$Sc$^+$/$^{48}$Ti$^+$ and \color{black}{9 }\color{black} of $^{48}$Ca$^+$/$^{48}$Ti$^+$, as well as 7 simultaneous cyclotron frequency measurements of $^{48}$Ca$^+$/$^{48}$Sc$^+$ as shown in Fig.~\ref{fig:Double_Res}. For the alternating cyclotron frequency data, the first ion that is measured is labeled ``ion $1$'' and the second is labeled ``ion $2$''. In total $N$ measurements are made of ion $2$ and $N + 1$ of ion $1$, so that each frequency measurement on ion $2$ performed at time $t_B$, $\nu_{c2}(t_B)$, is bracketed by two frequency measurements of ion $1$ performed at times $t_A$ and $t_C$, $\nu_{c1}(t_A)$ and $\nu_{c1}(t_C)$. A linear interpolation is performed on measurements $\nu_{c1}(t_A)$ and $\nu_{c1}(t_C)$ to obtain $\nu_{c1}(t_B)$, the cyclotron frequency of ion 1 at the time of the measurement on ion 2. The cyclotron frequency ratio $\nu_{c1}(t_B)$/$\nu_{c2}(t_B)$ is then obtained. 

\begin{figure}[b]
\includegraphics[width=0.9\columnwidth]{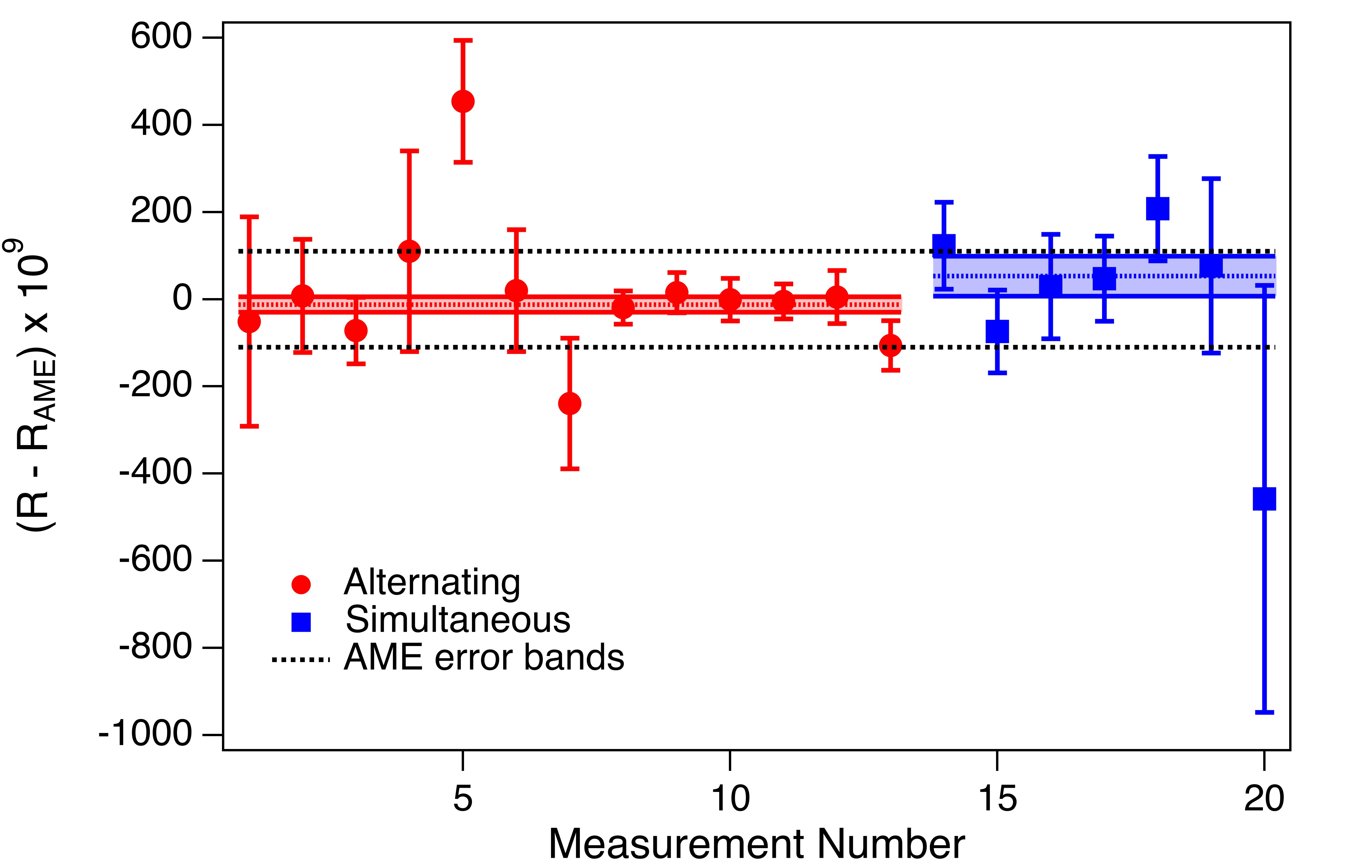}
\caption{(Color online) Individual measurements of the $^{48}$Ca$^{+}$/$^{48}$Sc$^{+}$ cyclotron frequency ratio obtained from alternating (red circles) and simultaneous (blue squares) resonance data (see text for details), compared to $R_{\textrm{AME}}$, the ratio obtained using AME data. The red (blue) dotted line and shaded band show the weighted average and error band for the alternating (simultaneous) data. The reduced-$\chi^2$'s for the fits are 1.5 (0.85). The black dashed lines show the error band for $R_{\textrm{AME}}$.}
\label{fig:48Ca_48Sc_Ratios}
\end{figure}

For the simultaneous cyclotron frequency measurements, a double fit was performed to both resonances in the TOF spectrum, and the quantities $\nu_{c1} = \nu_c(^{48}$Ca$^{+}$) and $\Delta \nu_c = \nu_{c2} - \nu_{c1}$ were obtained, where $\nu_{c2} = \nu_c(^{48}$Sc$^{+}$). The cyclotron frequency ratio was then determined as $R = 1 - \Delta\nu/\nu_1$.
Fig.~\ref{fig:48Ca_48Sc_Ratios} shows the individual cyclotron frequency ratios obtained from the alternating and simultaneous measurements of $^{48}$Ca$^{+}$ and $^{48}$Sc$^{+}$. The resulting average cyclotron frequency ratios for $^{48}$Ca$^{+}$/$^{48}$Sc$^{+}$, $^{48}$Sc$^{+}$/$^{48}$Ti$^{+}$, and $^{48}$Ca$^{+}$/$^{48}$Ti$^{+}$  are listed in Table~\ref{Tab:Ratios}.
%
\begin{table}[b]
\caption{\label{Table:Ratios} Cyclotron frequency ratios determined in this work. $N$ is the number of individual measurements of each ratio. Type refers to either alternating single resonance measurements (Alt.) or the simultaneous double resonances (Sim.) obtained for some of the $^{48}$Ca$^{+}$/$^{48}$Sc$^{+}$ data. $\sigma_{\textrm{stat}}$ is the statistical uncertainty in the weighted average, $\sigma_{\textrm{syst}}$ is the systematic uncertainty (see text) and $\bar{R}$ is the final average ratio with total uncertainty shown in parentheses.}
\begin{ruledtabular}
\begin{tabular}{lccccc}
Ion Pair & $N$ & Type & $\sigma_{\mathrm{stat}}$ & $\sigma_{\mathrm{syst}}$ & $\bar{R}$ \\
 &  &  & (ppb) & (ppb) & \\
\hline
(1) $^{48}$Ca$^{+}$/$^{48}$Ti$^{+}$  & 9 & Alt. & 19& 3.6 & 0.999 904 421(19) \\
(2) $^{48}$Sc$^{+}$/$^{48}$Ti$^{+}$  & 7 & Alt. & 15 & 3.6 &  0.999 910 692(15) \\
(3) $^{48}$Ca$^{+}$/$^{48}$Sc$^{+}$ & 13 & Alt. & 18 & 3.6 & 0.999 993 737(18) \\
   & 7 & Sim. & 45 & 3.6 & 0.999 993 805(45) \\
\end{tabular}
\end{ruledtabular}
\label{Tab:Ratios}
\end{table}
%
The statistical uncertainties listed in Table~\ref{Tab:Ratios} were obtained from the weighted average of the statistical uncertainties for each individual measurement. 

Possible systematic effects on the cyclotron frequency ratio were evaluated and included by adding them in quadrature to the statistical uncertainties. Potential contributions arising from magnetic field inhomogeneities, trap misalignment, higher-order distortions of the electrostatic potential, ion--ion interactions, temporal magnetic field variations, and relativistic effects were considered~\cite{Brodeur2012,Kakkar_PhD}.

In this work, pairs of ions used in a given cyclotron frequency ratio measurement have the same mass number ($A=48$) and charge state, resulting in a strong suppression of mass-dependent systematic shifts to the ratio. Relativistic shifts, for example, which depend on the ion mass and motional amplitude, have a completely negligible effect on the ratio due to the small relative mass difference between the pair of ions being evaluated. 

Systematic shifts due to imperfections in the trapping fields, described by higher-order terms in the expansion of the magnetic field and electrostatic potentials (see e.g. Ref.~\cite{Ketter2014}), as well as misalignment of the magnetic field axis, were investigated. This was done by varying the time at which the ions were captured in the trap after ejection from the cooler-buncher by 0.5~$\mu$s. This would affect the axial amplitude of the ions in the trap and, as a result, probe the effect of field imperfections. No statistically significant shift in the cyclotron frequency or derived ratio was observed under these different conditions, indicating that these effects are beyond the statistical sensitivity of the measurement.

Count-rate dependent shifts arising from ion--ion interactions were evaluated using a $z$-class analysis~\cite{Kellerbauer2003}. The data were grouped according to the number of detected ions per bunch, and independent fits, like in Fig.~\ref{fig:Single_Res}, were performed for each subset. No significant dependence of the cyclotron frequency on ion count was observed within uncertainties. Nevertheless, a maximum of 5 detected ions per shot was used in the final cyclotron frequency ratio measurement data.

Linear magnetic field drifts were accounted for by interleaving measurements of the two ions, as described above. The average magnetic field decay during the experiment was determined to be 1.2 $\times$ 10$^{-9}$ per h. The longest time interval between successive measurements of ion 1, which bracketed measurements on ion 2, was 3 hours. To be conservative we included a systematic uncertainty of 3.6 $\times$ 10$^{-9}$ to each cyclotron frequency ratio measurement as an upper limit estimate of non-linear magnetic field variations.

\section{Results and Discussion}

\subsection{Q values}

The cyclotron frequency ratios listed in Table~\ref{Tab:Ratios} can be converted into Q values, corresponding to the energy equivalent of the mass difference between parent and daughter atoms as follows,
\begin{equation}
    Q = (M_P - M_D)c^2 = (M_P - m_e)(1 - \bar{R}),
\label{Eqn:Q_value}
\end{equation}
where $M_P$ is the mass of the parent atom, $M_D$ is the mass of the daughter atom, $m_e$ is the mass of the electron~\cite{CODATA2022}, and $c$ is the speed of light. Atomic masses were obtained from AME2020~\cite{Wang_2021}. The uncertainty in $M_P$ contributes insignificantly to the uncertainty in $Q$ because $1 - \bar{R} < 10^{-4}$ for all ratios measured. In Eqn.~(\ref{Eqn:Q_value}) we have ignored electron binding energies for the singly-charged ions since these are $\sim$6 -- 7 eV for the isotopes we studied and are insignificant compared to the statistical uncertainties.

\begin{table}[b]
 \caption{\label{table:Q} Q value results for $^{48}$Ca $\beta$ and 2$\beta$-decays, and $^{48}$Sc $\beta$-decay obtained from the ratios listed in Table~\ref{Table:Ratios}, compared to AME2020 data. For $^{48}$Ca $\beta$-decay, the Q value is obtained from the alternating single resonance data (Alt.), the simultaneous double resonance data (Sim.), and from the difference between our $^{48}$Sc $\beta$-decay Q value and the $^{48}$Ca 2$\beta$-decay Q value listed in the AME2020~\cite{Wang_2021}.}
\begin{ruledtabular}
\begin{tabular}{cccccc}
\multirow{2}{*}{} & \multirow{2}{*}{Decay} & \multirow{2}{*}{Note} & \multicolumn{1}{c}{This work} & \multicolumn{1}{c}{AME2020}  \\
 & & & \multicolumn{1}{c}{(keV)} & \multicolumn{1}{c}{(keV)} \\
\hline
(1) & $^{48}$Ca -- $^{48}$Ti (2$\beta^-$) &  & 4269.23(87) & 4268.08(8)$^{*}$ \\
(2) & $^{48}$Sc -- $^{48}$Ti ($\beta^-$) &  & 3989.10(67) & 3988.9(4.9) \\
(3) & $^{48}$Ca -- $^{48}$Sc ($\beta^-$) & Alt. & 279.73(79) &   \\
 & & Sim. & 276.7(2.0) &  \\
 & & (1)$^{*}$ -- (2) & 278.98(67) &  \\
 & & Final Avg & 279.14(50) & 279.2(4.9)   \\

\end{tabular}
\end{ruledtabular}
\end{table}

\begin{figure}[t]
\includegraphics[width=0.9\columnwidth]{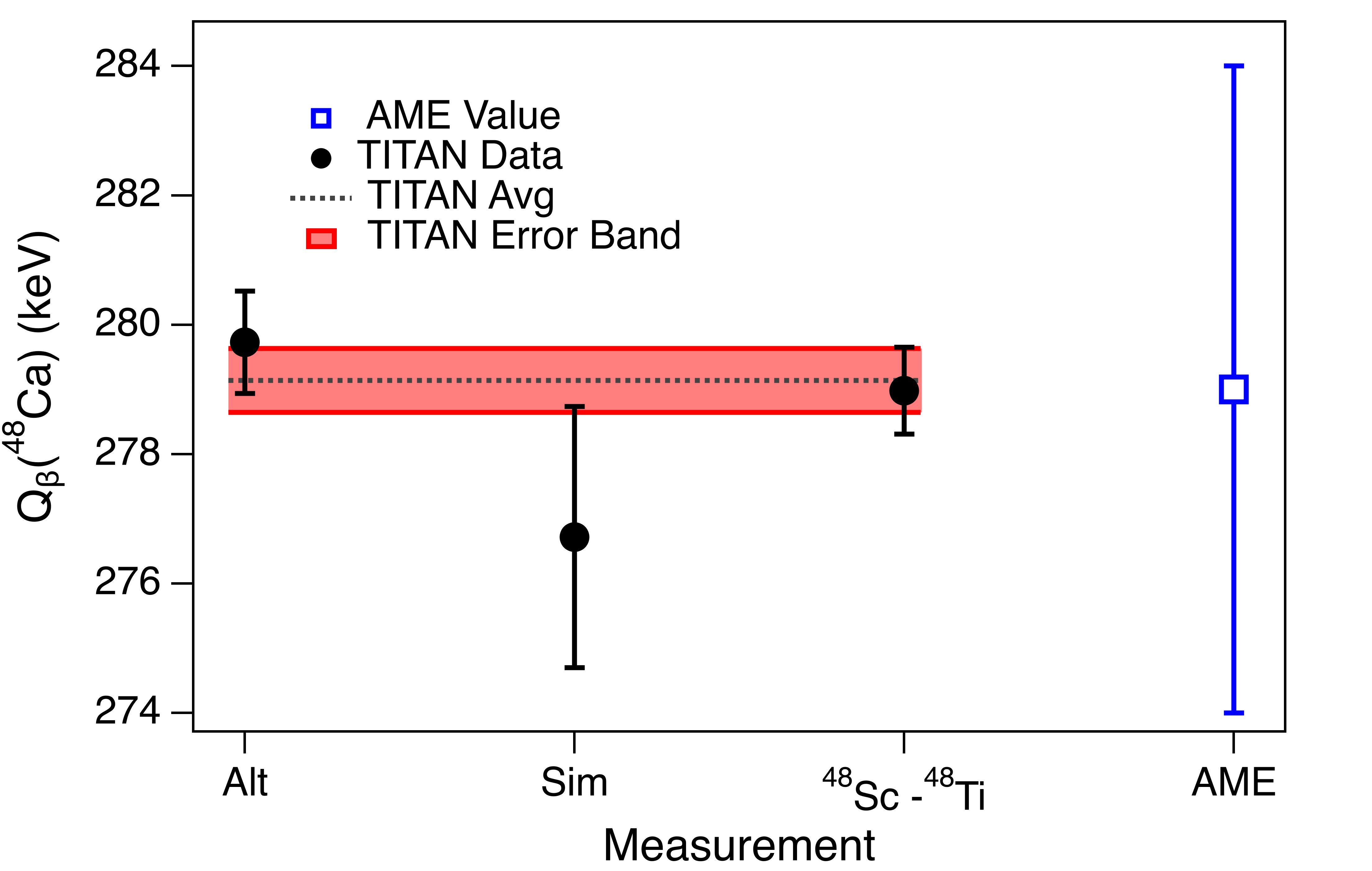}
\caption{(Color online) Q value measurements for $^{48}$Ca single $\beta$-decay obtained from the alternating (Alt) and simultaneous (Sim) $^{48}$Ca$^{+}$/$^{48}$Sc$^{+}$ measurements, and from the $^{48}$Sc$^{+}$/$^{48}$Ti$^{+}$ measurement combined with the precisely determined $^{48}$Ca -- $^{48}$Ti 2$\beta$-decay Q value from AME2020~\cite{Wang_2021}, as listed in Table~\ref{table:Q}. \label{fig: 48Sc_Q-values}} 
\end{figure}

The resulting Q values are listed in Table~\ref{table:Q}. Our measurement of the $^{48}$Ca double $\beta$-decay Q value is in agreement with the AME2020 value, which was obtained from Penning trap measurements~\cite{Redshaw2012,Bustabad2013,Kwiatkowski2012_48Ca_bb}, differing by 1.14(87) keV or 1.3$\sigma$. Our results for the $^{48}$Sc and $^{48}$Ca single $\beta$-decay Q values are in agreement with the AME2020 value, which is dominated by the 5 keV/$c^2$ uncertainty in the mass of $^{48}$Sc. 
Our $^{48}$Sc Q value is a factor of 7 more precise than AME value. Our two $^{48}$Ca single $\beta$-decay Q value results---one from the alternating data and one from the simultaneous data---listed as Alt. and Sim. in Table~\ref{table:Q} are, respectively, factors of 6 and 2.5 more precise than the AME value. 
In addition to our direct $^{48}$Ca single $\beta$-decay Q value measurement, we can also combine our result for the $^{48}$Sc $\beta$-decay Q value with the precisely known $^{48}$Ca 2$\beta$-decay Q value to obtain an additional independent determination of $Q_{\beta}$($^{48}$Ca). This result is listed in the second to last row of Table~\ref{table:Q}. These three results are plotted in Fig.~\ref{fig: 48Sc_Q-values}, and can be seen to be in good agreement within their uncertainties. We took a weighted average to obtain our final $^{48}$Ca single $\beta$-decay Q value: $Q_{\beta}$($^{48}$Ca) =  279.14(50) keV, which is a factor of 10 more precise than the AME2020 value.

By combining our final $^{48}$Ca single $\beta$-decay Q value result with the precisely determined $^{48}$Ca mass~\cite{Wang_2021,Kohler2016}, we also obtain a new value for the $^{48}$Sc mass excess: $ME$($^{48}$Sc) = -44\,504.01(50) keV/$c^2$. This result is in agreement with the AME2020 value, but is a factor of 10 more precise.

\subsection{Half-life}
We have performed a new shell-model study of $^{48}\text{Ca}$ single $\beta$-decay, using the GXPF1A interaction and the theoretical formalism of Refs.~\cite{Haaranen2014,Mustonen2006} that revises the theoretical single $\beta$-decay half-life based on the new Q value measurement of $Q_\beta = {279.14(50)}$ keV reported in this work. The updated half-life is:
\[
T_{1/2}^{\beta} = 5.09^{+0.05}_{-0.05} \times 10^{20} (g_A^{-2})\,  \text{y}  
\]
as compared with the older value of 
\[
T_{1/2}^{\beta} = 5.3^{+1.7}_{-1.2} \times 10^{20} (g_A^{-2})\,  \text{y} 
\]
reported in Ref. \cite{Haaranen2014} based on an older Q value of 278(5) keV~\cite{Audi2012_AME12}. We emphasize the significant improvement of about a factor of 30 in the uncertainty of the half-life due to the 10-fold improvement in the experimental uncertainty of the Q value. Here $g_A$ is the axial weak coupling constant, which is 1.27 in vacuum, but is quenched in nuclei. The quenched value in this case can be extracted by comparing the measured and calculated $^{48}$Ca 2$\nu\beta\beta$ decay half-life, which gives a result of 0.91(4), where the uncertainty is propagated from the $^{48}$Ca 2$\nu\beta\beta$ decay half-life result~\cite{Barabash2020}, and the same effective Hamiltonian as for the $^{48}\text{Ca}$ single $\beta$-decay half-life calculation was used.

This new half-life value, derived from the dominant 4$^{\mathrm{th}}$ forbidden unique transition, is still a factor of $\sim$5 longer than the current experimental lower limits, but is no longer influenced by uncertainty in the Q value. 
A recent compilation of all 2$\nu\beta\beta$ decay lifetimes determined $T_{1/2}^{2\nu\beta\beta}$($^{48}$Ca) = 5.96$^{+1.39}_{-1.08}\times 10^{19}$ y~\cite{Pritychenko2025}. Hence, the theoretical single $\beta$-decay to experimental double $\beta$-decay ratio is a factor of just 10, which  highlights the significance of $^{48}\text{Ca}$ as a potential benchmark for nuclear structure and $\beta\beta$-decay studies.

\section{Conclusion}

Using Penning trap mass spectrometry, we have performed the first direct measurement of the $^{48}$Ca single $\beta$-decay Q value. 
Our result confirms the previous AME value, which relied on the mass of $^{48}$Sc determined from nuclear reaction data, but is a factor of 10 more precise.

Using this new result we have performed updated shell model calculations to determine the $^{48}$Ca single-$\beta$-decay half-life. The uncertainty in the half-life has been reduced by a factor of about 30 compared to the result obtained using the previous Q value, providing an accurate and precise result that can be used for planning future experiments that could search for $^{48}$Ca single $\beta$-decay. Our result confirms that the single $\beta$-decay half-life is only a factor of about ten longer than that of the already observed double $\beta$-decay half-life, and about five times longer than the current experimental lower limit on the $^{48}$Ca single $\beta$-decay half-life. Hence, detection of $^{48}$Ca single $\beta$-decay could be within reach of future experimental efforts.

Finally, we have also performed the first direct measurement of the $^{48}$Sc single $\beta$-decay Q value, providing a direct experimental determination of the mass of $^{48}$Sc with 10-fold improvement in precision compared to the previous AME value.

\section*{Acknowledgments}
TITAN is funded by the Natural Sciences and Engineering
Research Council (NSERC) of Canada and through TRIUMF by the National Research Council (NRC) of Canada.
This research was supported by Central Michigan University, and the U.S. Department of Energy, Office of Science, Office of Nuclear Physics under Contracts No. DE-SC0022538 and DE-AC02-05CH11231. S.K. acknowledges support from the GETS/SEGS programs at the University of Manitoba. MR and DP ackowledge support from Central Michigan University.

\bibliography{48Sc_paper.bib} 

\end{document}